\documentstyle[12pt]{article}

\newcommand{\beq}{\begin{equation}}
\newcommand{\eeq}{\end{equation}}
\newcommand{\beqa}{\begin{eqnarray}}
\newcommand{\eeqa}{\end{eqnarray}}
\newcommand{\beqar}{\begin{eqnarray*}}
\newcommand{\eeqar}{\end{eqnarray*}}

\newcommand{\ie}{{\it i.e.,}\ }

\begin{document}

\begin{titlepage}

\vspace{.5in}
\thispagestyle{empty}

\begin{flushright}
TAUP 2141/94\\
gr-qc/9403027
March 1994\\
\end{flushright}
\vspace{.5in}

\begin{center}
{\bf \Large Thermodynamics and Evaporation of the 2+1-D Black
Hole}\\

\vspace{.4in}

B{\sc enni} R{\sc eznik}\footnote{\it e-mail: reznik@taunivm.tau.ac.il}\\

\medskip

{\small\it School of Physics and Astronomy} \\
{\small \it Beverly and Raymond Sackler Faculty of Exact Sciences}\\
{\small \it Tel Aviv University, Tel-Aviv 69978, Israel.}
\end{center}

\vspace{.5in}
\begin{center}
\begin{minipage}{5in}
\begin{center}
{\large\bf Abstract}
\end{center}
{\small
The properties of canonical and microcanonical ensembles of a
black hole with thermal radiation and the problem of black hole
evaporation in 3-D are studied.  In 3-D Einstein-anti-de Sitter
gravity we have two relevant mass scales, $m_c=1/G$, and
$m_p=(\hbar^2\Lambda/G)^{1/3}$, which are particularly relevant
for the evaporation problem. It is argued that in the `weak coupling'
regime $\Lambda<(\hbar G)^{-2}$, the end point of an evaporating
black hole formed with an initial mass $m_0>m_p$, is likely to be
a stable remnant in  equilibrium with thermal radiation.
The relevance of these results for  the information problem and
for the issue of back reaction is discussed. In the `strong coupling'
regime, $\Lambda>(\hbar G)^{-2}$ a full fledged quantum gravity
treatment is required. Since the total energy of thermal states in
anti-de Sitter space with reflective boundary conditions at spatial
infinity is bounded and conserved, the canonical and microcanonical
ensembles are well defined. For a given temperature or energy
black hole states are locally stable. In the weak coupling regime
black hole states are more probable then pure radiation states.
}
\end{minipage}
\end{center}
\end{titlepage}
\addtocounter{footnote}{-1}


\section{Introduction}

Three dimensional Einstein's gravity has no local degrees of
freedom and also no long range Newtonian-like interaction.
Local spacetime  curvature exits only in the presence of a local
matter source.\cite{2+1}
Notwithstanding this apparent
simplicity of the theory,
 Ba\~nados Teitelboim and Zanelli\cite{btz}  have
managed,
by adding
a negative cosmological constant source, and by a particular
identification of points in anti-de
Sitter spacetime,
to find a black hole solution that resembles
 in many features the 4-D
black hole. Their solution
is locally indistinguishable from
anti-de Sitter spacetime. The curvature is constant, but globally
the solution
 has the topology of a black hole.\footnote{For
further study of the black hole see references $[3-8]$.}

The metric of a 3-D black hole, of
mass $m$ and vanishing angular momentum and charge, in a static
coordinate system is\footnote
{In the literature Newton's constant
is sometimes taken as $G=1/8$.
We chose the usual action $1/16\pi G\int R$.}
\beq
ds^2 = \bigg(\Lambda r^2-8Gm\bigg)dt^2 - \bigg(\Lambda r^2-
8Gm\bigg)^{-1}dr^2 - r^2d\theta^2
\label{bh}
\eeq
where $\Lambda$ is the negative of the cosmological constant, and the
location of the horizon is given by $r_h=\sqrt{8Gm/\Lambda}$.

The entropy of the black hole is given by $S=
L/4\hbar G +C_d$ ($L=2\pi r_h$), where $C_d$ is an unknown additive
constant.\cite{entropy} It can be shown that for any classical
evolution
this entropy satisfies $\delta S\ge 0$, provided that
suitable
energy positivity conditions hold.\cite{reznik92} The entropy is
functionally related
to the Hawking temperature\cite{hawking} of the black hole $T_H
=\hbar\kappa/2\pi$
($\kappa=\Lambda r_h$ is the surface gravity on evaluated on the
horizon), via $dm=TdS$,
in a complete analogy to the
first law of thermodynamic. Therefore, the 3-D black hole satisfies
all the usual mechanical laws analogous to the laws of
thermodynamics.\cite{mechanical,reznik92}

In spite of these similarities, the 3-D black hole differs
significantly from the ordinary 4-D Schwarzschild black hole.
We note that the Hawking temperature decreases like $m^{1/2}$ as the
black hole
evaporates. Naively, by Stefan's law,
 the time required for a complete
evaporation is therefore
infinite. However at the `end point' the spacetime
metric does not yield empty anti-de Sitter spacetime but rather
a throat of zero radius that can be regarded as a sort of
`extremal' black hole.\cite{btz} The empty anti-de Sitter
spacetime is separated
by a negative energy gap from the spectra of black hole states.
Furthermore,
the metric of the black hole is asymptotically  anti-de Sitter,
and spatial infinity
is time-like. Therefore, as is well known, spacetime is not globally
hyperbolic\cite{haw&ell} and one has to impose reflective boundary
conditions at infinity in order to have a well defined Cauchy
problem.\cite{ads}  As we shall show,
these features significantly modify the thermodynamics
of the black holes and the problem of the final state.

In this work we study the
canonical and microcanonical ensembles of a black hole and thermal
radiation and the problem of black hole evaporation in 3-D.
The total energy of a thermal state in anti-de Sitter spacetime
is bounded, and with reflective boundary conditions
at spatial infinity it is also  conserved.  Therefore,
the  canonical partition function $Z(\beta)$
and the microcanonical density function $N(E)$ are well defined.
Some aspects of the problem resembles those of a
4-D Schwarzschild black hole in an anti-de Sitter
 spacetime.\cite{haw&pag}
However, contrary to the 4-D case,
a black hole of mass $m>(\hbar^2\Lambda/G)^{1/3}$ is always
in local equilibrium with a thermal radiation state.
When tunneling between states is also considered, the value of the
cosmological constant determines whether a black
hole with radiation or pure radiation without a black hole,
is more probable and  determines the final state
of the system.

Contrary
to 4-D gravity, in 3-D Einstein-anti-de Sitter gravity we have
two
relevant mass scales, $1/G$ and $(\hbar^2
\Lambda/G)^{1/3}$, that dictate the physical content of of the
model.
The relation of these two scales is determined by the value of the
cosmological constant. Only for $\Lambda=(G\hbar)^{-2}$ do
the two scales coincide.
The existence of two mass scales is particularly relevant for
the final state of an evaporating black
hole.  It is argued that for  $\Lambda<(\hbar G)^{-2}$ the end
point of a black hole formed with an initial
mass $m_0>(\hbar^2\Lambda/G)^{1/3}$ is likely to be a stable remnant.
For the particular case
$m_0>1/G>(\hbar^2\Lambda/G)^{1/3}$ the system will wind up
as a black hole of mass $m_{bh} (m_0)$ in a state of
equilibrium
with thermal radiation of total energy $m_{rad}=m_0 - m_{bh}$.
When the mass of the initially formed  black hole $1/G>m>(\hbar^2
\Lambda/G)^{1/3}$,
it is argued
that the black hole will radiate at a low rate, and
reach  gradually equilibrium. For $m<(\hbar^2\Lambda/G)^{1/3}$ the
semi-classical approximation breaks down.
In the regime of  `large curvature' $\Lambda>(\hbar
G)^{-2}$ things are more complicated, a full fledged
quantum gravity treatment is required .

The paper proceeds as follows. In section
2. we discuss the basic
units which are relevant to the problem of a black hole in anti-de
Sitter spacetime. The
canonical and microcanonical ensembles for a black hole and radiation.
are studied in sections 3. and 4 respectively.
The problem of black hole
evaporation is examined  in section 5. Finally, we discuss
the relevance of our results to the information problem and to the
issue of back reaction in section 6.
Throughout the paper we adopt the units $k_B=c=1$.

\section{Basic Units}

In 4-dimensions, the Planck mass provides the basic scale of the
theory. As the mass of the Schwarzschild black hole reaches the
Planck scale, the fluctuations of the geometry,
and the back reaction of the emitted radiation,  becomes large.
The Hawking temperature also
reaches the Planck scale.

The situation is different for the 2+1 black hole.
One can not
construct a basic Planckanian mass scale out of Newton's constants
$G$ and $\hbar$ alone
since $G$  has a dimension of $(mass)^{-1}$. The cosmological
constant (of dimension $(length)^{-2}$) is needed to
fix the radius of the black hole's horizon.
Combining $\Lambda$, $G$ and $\hbar$ yields various possible
definitions of a basic mass unit.

On physical grounds, we note that the theory singles out two
basic mass units. The fluctuations of the black hole
geometry become important when $r_{horizon}\sim r_{compton}$, \ie
the radius of the black hole becomes comparable to the Compton wave
length.  This yields a 3-D
analog of the Planck mass
\beq
m_p^{(3)} = \Biggl( {\hbar^2\Lambda\over G} \Biggr)^{1/3}\equiv m_p.
\label{mplack}
\eeq
However, the Hawking temperature (or the wavelength of the radiation)
of a black hole with mass $m_p$ does not coincide in general
with this `Planck scale'.

Newton's constant yields a second basic mass unit
\beq
m_c = {1\over G}
\eeq
The physical significance of this `classical' mass unit is
 that
for $m>m_c$ ($m<m_c$) the wave length $\lambda$ of the Hawking
radiation satisfies $\lambda<r_h$ ($\lambda>r_h$).
The relation of the two mass units is determined by the value of the
cosmological constant. For $\Lambda=(\hbar G)^{-2}$, $m_c=m_p$.
Notice that in the  `strong coupling' regime $\Lambda>
(\hbar G)^{-2}$, the classical action
${1\over 8\pi G}\int(R+2\Lambda)\sqrt{g}d^3x$ of anti-de Sitter
spacetime is smaller than  the quantum
action $\hbar$. Therefore, it  is likely that quantum fluctuations
are significant and classical considerations may not be adequate.

\section{The Canonical Ensemble}

Thermal states in anti-de Sitter space can be constructed by a
periodic identification of the Euclidean time coordinate $\tau_e=it$
with a period $\beta=T^{-1}$.\cite{4d-ads} The thermal radiation will
be in
equilibrium, with local temperature $T/\sqrt{g_{tt}}$, in the static
coordinate system (\ref{bh}),
and hence its presence breaks the
$SO(2,2)$ invariance of the `empty' anti-de Sitter spacetime
to $SO(2)\times SO(2)$.  Due to the red shift with respect to the
preferred origin, the local temperature decreases as $1/r$ at infinity
and the local energy is expected to decrease like $1/r^3$.
The total energy of the thermal radiation is therefore bounded.
Neglecting the back-reaction of the radiation on the geometry the
local energy of a conformally coupled scalar field in a thermal state
 can be approximated by\cite{haw&pag,4d-ads}
\beq
T^t_t \simeq  {a_0\over\hbar^2} {T^3\over {g_{tt}(r)}^{3/2}} + O(T^2)
\eeq
$a_0$ is a dimensionless Stefan-Boltzmann
constant.
The total energy of the state is approximated by
\beq
E_{rad} \simeq {a_o\over\hbar^2\Lambda} T^3 + O(T^2).
\eeq
The entropy and free energy of the radiation can be easily derived,
\beq
S_{rad} \simeq  {3a_0\over2} {1\over\Lambda\hbar^2} T^2+O(T),
\label{t-entropy}
\eeq
and
\beq
F_{rad} \simeq -{a_0\over2} {1\over\Lambda\hbar^2} T^3 + O(T^2).
\label{frad}
\eeq
The partition function is given by
\beq
Z_{rad}(\beta) \simeq\exp\biggl({{a_0\over2} {1\over\Lambda\hbar^2}
\beta^{-2} + O(\beta^{-1})}\biggr).
\label{zrad}
\eeq
A state of pure thermal radiation is unstable against collapse to a
black at sufficiently high temperature.
This can be seen by noting that for $r<<\Lambda^{-1/2}$ the total
energy of the radiation in a sphere of radius $r$ increases as
$r^2T^3$, while for the 3-D black hole we have $m_{bh} \sim r^2$.
This yields an upper bound $T_{critical}$ on the maximal value of the
temperature,
\beq
T<T_{critical} = \Biggl({\Lambda\hbar^2\over G}\Biggr)^{1/3}= m_p,
\label{tcritic}
\eeq
which corresponds to a state of total energy $E_{rad}<m_c$.
Alternatively, we may
look at a solution of Einstein's equation with a negative
cosmological
constant and a spherically symmetric source. We have\cite{entropy}
\beq
g^{rr}(r) = \Lambda r^2 - 16\pi G\int_0^r \rho r dr + c.
\eeq
For $c$=1 and $\rho=0$ the metric corresponds to an
anti-de Sitter space.
With $\rho=a_0T^3/\hbar^2$ we note that a horizon forms unless
(\ref{tcritic}) is maintained.

We now examine states that contain a black hole.
The Hawking temperature of a black hole of mass $m$ is $\hbar\Lambda
r_h(m)/2\pi$, therefore
\beq
m(T) = \Biggl({2\pi\over\hbar}\Biggr)^2{1\over\Lambda G} T^2.
\eeq
This semi-classical approximation
safely holds only when $m>m_p$, this yields
\beq
T> (\hbar^4\Lambda^2 G)^{1/3} \equiv T_1
\label{validity}
\eeq
The heat capacity of the black holes ${\partial m/\partial T}$
is always positive. Therefore,
for  a given temperature $T$ there exists a
 black hole of mass $m(T)$ in a state of local
equilibrium with the radiation.

Contrary to a Schwarzschild black hole in an asymptotically flat space
the canonical partition function is well defined.
The entropy of the black hole is given by
\beq
S_{bh} = {\pi\over2\hbar G} r_h + C_d =
{\pi\over2\hbar}\Biggl({8\over\Lambda G}\Biggr)^{1/2}m^{1/2}  +C_d
\label{bh-entropy}
\eeq
Assuming the integration constant $C_d$ is a finite
number\cite{entropy}, the density of states of the black holes is
\beq
N(m) \sim e^S \sim e^{m^{1/2}}.
\eeq
The integral that defines the partition function
\beq
Z(\beta) = \int_0^\infty N(m) e^{-m\beta} dm
\label{z}
\eeq
therefore converges.

$Z(\beta)$ can be computed directly from (\ref{z}) or from the
the Euclidean path integral approach by a saddle point
approximation.\cite{continued} In either of the two ways we find
\beq
\log Z = - I_e = {F\over T}
        = \Biggl({2\pi\over\hbar^2}\Biggr)^2 {\beta^{-1}\over
            8G\Lambda}
\label{zbh}
\eeq
and the free energy
\beq
F_{bh} = -\Biggl( {2\pi\over\hbar^2} \Biggr)^2 {T^2\over8G\Lambda}.
\label{fbh}
\eeq
Comparing (\ref{frad}) and (\ref{fbh}) we note that for
$T>m_c$, $F_{rad} < F_{bh}$. In this case an initial state with
a black hole
will ultimately tunnel to a configuration with
pure radiation. The properties of the ensemble  depend on the value of
$\Lambda$.  For $\Lambda<(\hbar G)^{-2}$, we have $m_p< m_c$. Since
a pure state is unstable for $T>m_p$, in this case only black hole
configurations are stable.
Note however that by (\ref{validity})
our semi-classical consideration strictly hold only for
$T>T_1>m_c$.

On the other hand, in the strong coupling regime, $\Lambda>(\hbar
G)^{-2}$, we have $m_p>m_c$. If the large fluctuations of the
metric can be disregarded, then
for $T_1<T<m_p$ ($T_1<m_c$) a configuration with pure thermal radiation
is stable, and for $T>m_p$ only black hole states are stable. For
$T<T_1$ (which corresponds to $m<m_p$) our semi-classical
approximation for the black hole breaks.

\section{The Microcanonical Ensemble}

The boundary conditions at infinity\cite{ads} insure that all the
outgoing flux
to infinity is reflected back and the total energy is conserved.
Therefore, one can consider the microcanonical ensemble and evaluate
the number of states $N(E)dE$
of the system between $E$ to $E+dE$.
Given the canonical partition function $Z(\beta)$, $N(E)$ is
expressed by the inverse Laplace transform
\beq
N(E) = \int^{+i\infty}_{-i\infty} Z(\beta)e^{\beta E}d\beta.
\label{ne}
\eeq
For $T<T_c$ the partition function of a thermal state is given by
(\ref{zrad}).  The integral
(\ref{ne}) has a saddle point at
\beq
\beta \simeq \biggl( a_0 {1\over \Lambda\hbar^2}\biggr)^{1/3}
E^{-1/3}
\eeq
and in the stationary phase approximation $N(E)$ is given by
\beq
N_{rad}(E)
\simeq \exp\Biggl( {3\over2}
 \biggl({a_0\over\Lambda\hbar^2} \biggr)^{1/3}
E^{2/3} \Biggr)
\label{nrad}
\eeq
This expression for $N(E)$ holds for $E<m_c$, which corresponds
to a saddle point at $\beta > m_p^{-1}$.

{}From the partition function of the black hole (\ref{zbh})
we find a saddle point at
\beq
\beta = \biggl( {2\pi\over\hbar} \biggr)
        \biggl( {1\over8G\Lambda} \biggr)^{1/2} E^{-1/2}
\eeq
and
\beq
N_{bh}(E) \simeq \exp \Biggl(
      {4\pi\over\hbar} \Bigl(8G\Lambda\Bigr)^{-1/2} E^{1/2} \Biggr)
\label{nbh}
\eeq
Equation (\ref{nbh}) holds provided that $E>m_p$.
Comparing equations (\ref{nrad}) and (\ref{nbh}) we find that
 for $E>E_1=G^{-3}\Lambda^{-1}\hbar^{-2}$, $N_{bh}<N_{rad}$.
$E_1$ corresponds to the total energy of a thermal
distribution with temperature $m_c$.
For $\Lambda<(G\hbar)^{-2}$ $E_1>m_c$ and hence black hole states in
equilibrium
are always more probable. Any configuration with initial energy
$E>m_p$ will settle finally to a state of a black hole in equilibrium
with thermal radiation.
At the stationary point, the temperature of the system
will be determined by the equation
\beq
E \simeq  {a_0\over\hbar^2\Lambda}\beta^{-3}
 + \biggl( {2\pi\over\hbar}
  \biggr)^2 {1\over\Lambda G} \beta^{-2}
\label{aqu}
\eeq
Since  $E, T > 0$,  (\ref{aqu}) has only one solution for $\beta$.

For  $\Lambda>(G\hbar)^{-2}$, and $m>m_p$ black hole states are more
probable. Thermal radiation states are more likely at $E_1<E<m_c$, but
in this regime our semi-classical approximation breaks.

\section{Black Hole Evaporation}

We have seen in the previous section that a configuration of given
total energy $m_0>m_p$ evolves finally to  a state of a black
hole in equilibrium with thermal radiation.
This suggests that even in a dynamical situation of matter
collapsing to a black hole the final state may be a black hole of
a mass $m(m_0)$.

Coming to the dynamical problem of black hole evaporation it
is essential to ask whether the black hole can really lose mass.
During the evaporation
the geometry of space remains
asymptotically anti-de Sitter.
Any massive particle emitted
by the black hole that travels along a geodesic path
will not reach spatial infinity. Rather it will
be captured\cite{geodesic}
back by the black hole on a time scale of $\Lambda^{-1/2}$.
Massless particles do  reach spatial
infinity. However, since anti-de
Sitter spacetime is not globally hyperbolic, one must impose
appropriate
boundary conditions at spatial infinity in order to get a well
defined Cauchy problem. This can be obtained by imposing either
von-Neumann or Dirichlet boundary conditions on the matter
fields\cite{ads} As a result
energy is conserved and the energy flux
is reflected back from infinity. Therefore, even
massless particles return to the black hole on the same time scale.
(Since space is not empty
the motion for both cases, $m=0$ and $m\ne0$,
 will not necessary be geodesic
but this should not alter this basic picture.)

Let a  matter distribution of total mass $m_0$
in an anti-de Sitter space collapse and form a black hole.\cite{mann}
We first consider the possibility that $\Lambda<(G\hbar)^{-2}$, \ie
$m_c>m_p$.
If the initial mass satisfies $m_0>m_c$ then the typical wave length
related to Hawking's temperature is smaller than the radius of the
black hole. Therefore, the process of
energy emission can be well approximated
since $\lambda<r_h$
by Stefan's radiation law,
at least until the black hole's mass reduces to $m \sim m_c$.
A simple integration yields the result
\beq
\Delta t = {1\over\hbar G^2 \Lambda}\biggl(m_c^{-1}-m_0^{-1}\biggr)
>\Lambda^{-1/2}.
\eeq
Hence the emitted radiation has sufficient time to reach back the black
hole. Since the whole process takes place for black hole
much above the Planck scale
this result strongly indicates that the final state is that of a
macroscopic black hole in a state of equilibrium with thermal
radiation at a temperature given by Eq. (\ref{aqu}).
When the initial mass is $m_p<m<m_c$ the situation is drastically
altered. The wavelength of Hawking's radiation $\lambda$ satisfies
$\lambda>r_h$. We argue that the black hole can not emit radiation
at such a long wavelength. Since $\lambda\sim
\Lambda^{-1/2}$,  the reflected flux of a single wave  reaches
back to the black hole before it can get away
and  interferes destructively. In other words real particle can not
materialize. The
situation is very similar to that of an exited atom inside a cavity.
If the wavelength related to the energy gap between the
energy levels of the atom
is of the order of the size of the box or larger,
the probability to emit a photon will be significantly smaller.
Therefore,
in this case, we expect a much slower rate of emission, and the black
hole will have sufficient time to reach an equilibrium.

The state of affairs is more complicated for  $\Lambda>(G\hbar)^{-2}$.
As noted in section 2., a semi-classical description
of spacetime may not be appropriate for this `strong coupling'
regime. Nevertheless, let us examine the evaporation
assuming a classical spacetime background.
In this case the typical wavelength of
Hawking radiation will satisfy $\lambda<r_h$  all the way to the Planck
mass. Since $m_p>m_c$, the black hole will emit radiation,
and the rate of emission  can be described adequately up to the
Planck scale by Stefan's law. It is easily verified
that the time required for the black hole to reach the Planck mass
in this case satisfies
$\Delta t < \Lambda^{-1/2}$. Hence, the black hole may
lose its mass much before the arrival of the return flux. If
Stefan's law is extrapolated beyond $m_p$, $\Delta t \to
\infty$. However, in this case, without a quantum model for the black
hole, it is not clear at what rate the black
hole radiates if it does at all, or what is the cross section for
absorption.

\section{Discussion}

We have seen that in the small coupling regime, $\Lambda<(\hbar
G)^{-2}$, the end point of an evaporating 3-D black hole is likely
to be a macroscopic black hole whose mass is determined  by the
initial configuration.
Consequently, there should not be an  information
problem in this case.  In general
the entropy $S_0$ of the
initial state of collapsing matter with  mass $m_0>m_c$ can not
exceed
the entropy of thermal radiation with the same mass. We have
from eq. (\ref{t-entropy})
$S_0<S_{rad}\sim (\hbar^2\Lambda)^{-1/3}m_0^{2/3}>1$.
On the other hand, the entropy of the final black hole state is give
by eq. (\ref{bh-entropy}), and it is  easily verified
to be of the same order of magnitude of $S_0$ above
when the integration constant
$C_d$ is bounded and small.
Therefore, there is sufficient room for storing  the information
in the `surface' states alone without the introduction of a large
degeneracy constant $C_d$.
Our conclusion agrees with the suggestion
made in reference \cite{entropy} (where the relation of the entropy
to the process of tunneling from an initial matter configuration
to a  black hole was
studied), that the degeneracy of the 3-D black hole
is likely to be finite or zero, \ie $C_d\sim 0$, and that the
entropy is given simply  by the length of the horizon.
Our semi-classical arguments may not apply to the
strong coupling regime ($\Lambda>(\hbar G)^{-2}$).

It is also interesting to note that for the case
 $m>m_c$, $\lambda<r_h$, \ie
during the evaporation process
photons are emitted from a small region of the horizon.
As indicated in the following this fact may seem at first to be
paradoxical.
Having two scales of length, $\lambda$
and $r_h$, one may average the stress tensor
on an intermediate scale $l$, $\lambda<l<r_h$, and compute
its expectation value: $<\int T_{ab}>$.
The domains of size $l^2$ are
larger than the scale where
quantum fluctuations are important,
but still these domains are sufficiently close to the
horizon. Since also a large number of photons is emitted from
such a domain
in a time interval $l/c$, we could expect that
$<\int T_{ab}>$ behaves as a classical stress tensor.
However, a classical stress tensor should satisfies the
positivity requirement and hence, by the second law of black hole
mechanics,
the radius of the black hole's horizon can not decrease
and therefore, radiation can not be emitted.
On the other hand, we could compute the stress tensor on
small region of scale
$\lambda$
near the horizon and find negative energy,
and therefore conclude
that the black hole can radiate and lose mass.
Somehow the first  averaged, classically behaved,
stress tensor must also contain a negative contribution that accounts
for Hawking's radiation.
A similar `classical' limit does not exist in 4-D since
there the wavelength of the Hawking radiation is always of the
size of the horizon.
It is possible that by studying the
interpolation between the two cases, one could understand better
the problem of back reaction, at least in the 3-D context.

\vspace{.3in}
\noindent
{\bf Acknowledgments}\\
I would like to thank Yakir Aharonov and Aharon Casher for
very helpful discussions.
The research was supported in part by grant 425-91-1 of the
Basic Research Foundation, administered by the Israel Academy
of Sciences and Humanities .

\vfill
\eject

\end{document}